\documentclass[12pt,a4paper]{article}

\usepackage{amssymb}
\usepackage{epsf}

\begin{document}

\title{%
\Large \textbf{Spherical and non-spherical bubbles in cosmological
phase transitions}}
\author{\large Leonardo Leitao\thanks{%
E-mail address: lleitao@mdp.edu.ar}~  and Ariel M\'{e}gevand\thanks{%
Member of CONICET, Argentina. E-mail address: megevand@mdp.edu.ar} \\[0.5cm]
\normalsize \it IFIMAR (CONICET-UNMdP)\\
\normalsize \it Departamento de F\'{\i}sica, Facultad de Ciencias Exactas
y Naturales, \\
\normalsize \it UNMdP, De\'{a}n Funes 3350, (7600) Mar del Plata,
Argentina }
\date{}
\maketitle

\begin{abstract}
The cosmological remnants of a first-order phase transition generally
depend on the perturbations that the walls of expanding bubbles
originate in the plasma. Several of the formation mechanisms occur
when bubbles collide and lose their spherical symmetry. However,
spherical bubbles are often considered in the literature, in
particular for the calculation of gravitational waves. We study the
steady state motion of bubble walls  for different bubble symmetries.
Using the bag equation of state, we discuss the propagation of phase
transition fronts  as detonations and subsonic or supersonic
deflagrations. We consider the cases of spherical, cylindrical and
planar walls, and compare the energy transferred to bulk motions of
the relativistic fluid. We find that the different wall geometries
give similar perturbations of the plasma. For the case of planar
walls, we obtain analytical expressions for the kinetic energy in the
bulk motions. As an application, we discuss the generation of
gravitational waves.
\end{abstract}

\section{Introduction}

Cosmological phase transitions generically produce cosmic relics such
as topological defects \cite{vs94}, magnetic fields \cite{gr01},
baryon number asymmetries \cite{ckn93}, inhomogeneities
\cite{w84,h95}, and gravitational waves \cite{kt93,kkt94}. The walls
of expanding bubbles usually play a relevant role in the mechanisms
which generate these relics. Indeed, most of them depend on the
disturbance produced by the motion of bubble walls in the surrounding
plasma. Moreover, the wall velocity itself depends on such
hydrodynamics
\cite{s82,gkkm84,k85,mp89,ikkl94,kk86,l94,kl95,ms08,ms09} as well as
on the friction with the plasma \cite{fric}. The friction is
determined by microphysics, i.e.,  the interactions of particles with
the wall. The hydrodynamics is determined by the relativistic fluid
equations. The latent heat that is injected at the phase transition
fronts spreads out, causing reheating and bulk motions of the plasma.
There are essentially three propagation modes for the phase
transition front. A detonation, which is supersonic and is followed
by a rarefaction wave, a subsonic deflagration, which is preceded by
a shock front, and a supersonic deflagration, which is preceded by a
shock and followed by a rarefaction wave.

Several of the mechanisms which generate the aforementioned
cosmological relics operate when bubbles meet. For instance, in order
to produce gravity waves (GWs), bubbles must collide and lose their
spherical symmetry, since a spherical source cannot generate
gravitational radiation. The hydrodynamics of colliding bubble walls
is complicated. The fluid velocity and temperature profiles during
bubble collisions were studied using numerical simulations in Ref.
\cite{ikkl94}. In Ref. \cite{kk86}, a few configurations for the
collision of shock fronts were considered analytically. Both studies
were performed in 1+1 dimensions. In applications, further
simplifications are needed. For instance, in the case of
gravitational waves, the relevant quantity is the energy that is
injected into bulk motions of the plasma. This is parameterized by
the efficiency factor $\kappa$, which is defined as the ratio of the
kinetic energy in bulk motions to the released vacuum energy. In
calculations of $\kappa$, the interaction between bubbles is
neglected, and spherical bubbles are assumed. This treatment implies
two hypothesis, namely, that the motion of a bubble wall is not
affected by the perturbations other bubbles caused in the plasma, and
that the deformation of the bubble wall does not affect the energy
transfer from the wall to the plasma. The first assumption should be
correct for supersonic walls, which either do not have shock fronts
preceding them or the shock fronts are very close to the walls. For
subsonic walls, the main influence from other bubbles is the
reheating of the plasma. This effect must be taken into account in a
complete calculation of the phase transition \cite{h95,m01,m04}.
Regarding the dependence on the wall geometry, some progress can be
made by comparing different bubble symmetries. Of course, after a few
collisions bubbles will take arbitrary forms. Nevertheless,
considering a few specific symmetries will be useful to tell whether
the disturbance of the fluid has a strong dependence on the wall
geometry or not. Furthermore, it is particularly important to study
the case of planar walls, for which analytical results can be
obtained.

The growth of plane, cylindrical, and spherical bubbles
(equivalently, bubbles in 1, 2, and 3 spatial dimensions,
respectively) was considered in Ref. \cite{k85} for the case of
deflagrations. The fluid velocity profiles are quite different. In
particular, for planar walls the shock wave preceding the phase
transition front has a constant fluid velocity, whereas in the
cylindrical and spherical cases the fluid velocity falls quickly in
front of the wall. This is because in higher dimensions there is more
room for the shock wave to carry away the injected energy.
Nevertheless, the total amount of kinetic energy of the fluid must be
a fraction of the released latent heat (another fraction goes into
thermal energy), and there is no reason for this fraction (and for
the ratio $\kappa$) to have a strong dependence on the bubble wall
geometry. For planar walls, the velocity profile of the fluid can be
obtained analytically (see, e.g., \cite{k85}), and one expects to
find analytical formulas for quantities such as the efficiency factor
as well. An analytical approximation for $\kappa$ was obtained in
Ref. \cite{m08} for small wall velocities. Recently, the efficiency
factor was calculated numerically for spherical bubbles \cite{ekns10}
in the whole wall velocity range.

In this paper we address the issue of the effect of the wall geometry
on the disturbance caused by the motion of the walls. Thus, we study
the hydrodynamics of spherical, cylindrical, and planar bubble walls.
In particular, we show that the efficiency factor does not differ
significantly between the different wall geometries. For planar
walls, we obtain the efficiency factor analytically for any wall
velocity. We also discuss the generation of GWs in the electroweak
phase transition. The plan of the paper is the following. In the next
section we review the dynamics of the first-order phase transition.
In section \ref{fluid} we calculate the fluid profiles and the
efficiency factor for the three bubble symmetries. In section
\ref{planar} we study in detail the case of planar walls. We discuss
the different hydrodynamical modes and we derive analytical formulas
for $\kappa$. In section \ref{gw} we apply our results to the
estimation of the gravitational wave signal from the electroweak
phase transition. The conclusions are in section \ref{conclu}.

\section{Phase transition dynamics}

A cosmological phase transition occurs when the free energy of the
model (i.e., the finite-temperature effective potential) depends on
an order parameter $\phi $ (e.g., a Higgs field), such that the free
energy density $\mathcal{F}\left( \phi ,T\right) $ has a minimum
$\phi _{+}\left( T\right) $ at high temperatures, and a different
minimum $\phi _{-}\left( T\right) $ at low temperatures (we shall use
a ``$+$'' index for variables in the high-temperature phase and a
``$-$'' index for the low-temperature phase). For a first-order phase
transition, there is a temperature range in which the two minima
coexist and a barrier in the free energy separates them. The critical
temperature $T_c$ is that at which the two minima have the same free
energy. Below $T_c$, bubbles of the stable phase nucleate and grow.
The nucleated bubble is a configuration  $\phi =\phi \left(
r,t\right) $ with spherical symmetry, such that at the center of the
bubble the system is in the low-$T$ phase, whereas far from this
point the system is in the high-$T$ phase. Hence, we have $\phi
\left( r=0\right) =\phi _{-}\left( T\right) $ and $ \phi \left(
r=\infty \right) =\phi _{+}\left( T\right) $. There is a region, the
``bubble wall'', in which $\phi$ varies continuously from $\phi _-$
to $\phi _+$. In this work, we shall assume for simplicity an
infinitely thin wall separating the two phases.

We are interested in the energy in bulk motions of the fluid which is
caused by the moving walls. We assume that the plasma is a perfect
relativistic fluid with four-velocity field $u^{\mu }=
(\gamma,\gamma\mathbf{v})$, with $\gamma =1/\sqrt{1-v^{2}}$. The
energy-momentum tensor is of the form
\begin{equation}
T^{\mu \nu }=\left( e+p\right) u^{\mu }u^{\nu }-pg^{\mu \nu }, \label{tmunu}
\end{equation}
where $e$ and $p$ are the energy density and pressure in the proper
system of the fluid element  \cite{landau}.  The energy density is
given by $T^{00}=w\gamma ^{2}-p=\left( e+pv^{2}\right) \gamma ^{2}$,
and the kinetic energy density is defined as $
e_{\mathrm{kin}}=T^{00}\left( v\right) -T^{00}\left( 0\right) $.
Therefore, we have $e_{\mathrm{kin}}=T^{00}-e=wv^{2}\gamma ^{2}$.

All the thermodynamical quantities can be derived from the free
energy densities $\mathcal{F}_{+}\left( T\right) \equiv
\mathcal{F}\left( \phi _{+}\left( T\right) ,T\right) $ and
$\mathcal{F}_{-}\left( T\right) \equiv \mathcal{F}\left( \phi
_{-}\left( T\right) ,T\right) $  for each phase. Thus, the pressure
is given by $p=-\mathcal{F}$, the entropy density by $s=dp/dT$, the
energy density by $e=Ts-p$, and the enthalpy by $w=e+p=Ts$.  At the
critical temperature, the pressure in the two phases is the same,
i.e., $ p_{+}\left( T_{c}\right) =p_{-}\left( T_{c}\right) $.
However, other quantities such as the energy, entropy, and enthalpy
are different even at $T=T_{c}$. The latent heat is defined as the
energy density difference $L=\Delta e\left( T_{c}\right) =\Delta
w\left( T_{c}\right) =T_{c}\Delta s\left( T_{c}\right) $.

It is useful to consider a simplified model which exhibits the
general features of a phase transition. This allows in particular to
obtain results which depend on a few parameters. Then, the results
can be applied to realistic phase transitions by calculating these
parameters in specific models. Therefore, we shall use the bag
equation of state (EOS),
\begin{equation}
\begin{array}{ccc}
e_{+}=a_{+}T^{4}+\varepsilon , &  & e_{-}=a_{-}T^{4}, \\
p_{+}=\frac{1}{3}a_{+}T^{4}-\varepsilon &  & p_{-}=
\frac{1}{3}a_{-}T^{4},
\end{array}
\label{EOSbag}
\end{equation}
with $\varepsilon >0$ and $a_{+}>a_{-}>0$. This system has two
components, namely, a vacuum energy density $\varepsilon $ and a
radiation energy density $aT^{4}$. The condition $a_{+}>a_{-}$
implies that some of the radiation degrees of freedom disappear after
the phase transition. The vacuum energy density is positive in the
high-temperature phase and vanishes in the low-temperature one. In
this model, the critical temperature is given by $T_{c}=\left(
3\varepsilon /\Delta a\right) ^{1/4}$, where $\Delta a=a_{+}-a_{-}$,
and the latent heat is related to the vacuum energy density by
$L=4\varepsilon $. In both phases, the speed of sound is given by
\begin{equation}
c_{s}^{2}\equiv\partial p/\partial e =1/3.  \label{cs}
\end{equation}

We shall assume for simplicity a stationary state in which the wall
moves with a constant\footnote{In a real phase transition, even after
reaching the stationary state, the wall velocity may vary due to the
adiabatic cooling of the universe and the release of latent heat
(see, e.g., \cite{h95,m01,m00}).} velocity $v_{w}$. We aim to
consider three kinds of bubble wall geometry, namely, a spherical, a
cylindrical and a plane wall. In this approximation, at time $t$ the
wall is at a distance $R_{b}=v_{w}t$ from a point, axis, or plane,
and the volume of the bubble is of the form $
V_{b}=c_{j}R_{b}^{j+1}/\left( j+1\right) $, where $j=2,1$, or $0$ for
the spherical, cylindrical or planar case, respectively (the factor
$c_{j}$ will cancel out in our calculations). The kinetic energy in
bulk motions of the fluid is given by $E_{kin}=c_{j}\int_{0}^{\infty
}wv^{2}\gamma ^{2}R^{j}dR$. The {efficiency factor} is defined as the
ratio of the kinetic energy to the released vacuum energy,
\begin{equation}
\kappa =E_{kin}/\left( \varepsilon V_{b}\right) .
\end{equation}
Thus, we can write
\begin{equation}
\kappa =\frac{j+1}{\varepsilon \xi _{w}^{j+1}}\int_{0}^{\infty }wv^{2}\gamma
^{2}\xi ^{j}d\xi ,  \label{kappa}
\end{equation}
where $\xi =R/t$ and $\xi _{w}=R_{b}/t=v_{w}$.

\section{Fluid profiles and kinetic energy} \label{fluid}

\subsection{The fluid equations}

The fluid equations are obtained from the conservation of the
energy-momentum tensor (\ref{tmunu}),
\begin{equation}
\partial _{\mu }T^{\mu \nu }=0.  \label{consT}
\end{equation}
If we denote by $r$ the distance from the symmetry point, axis or plane, and
$t$ the time from nucleation, Eqs. (\ref{consT}) become \cite{k85}
\begin{eqnarray}
\partial _{t}\left[ \left( e+pv^{2}\right) \gamma ^{2}\right] +\partial _{r}
\left[ \left( e+p\right) \gamma ^{2}v\right] &=&-\frac{j}{r}\left[ \left(
e+p\right) \gamma ^{2}v\right] ,  \nonumber \\
\partial _{t}\left[ \left( e+p\right) \gamma ^{2}v\right] +\partial _{r}
\left[ \left( ev^{2}+p\right) \gamma ^{2}\right] &=&-\frac{j}{r}\left[
\left( e+p\right) \gamma ^{2}v^{2}\right] . \label{fluid0}
\end{eqnarray}
Since there is no characteristic distance scale in the problem, it is
usual to assume the \emph{similarity condition}, namely, that $e,p$
and $v$ depend only on the combination $\xi =r/t$. Thus, we have
\begin{eqnarray}
\left( \xi -v\right) \frac{e^{\prime }}{w} &=&j\frac{v}{\xi }+\gamma
^{2}\left( 1-v\xi \right) v^{\prime },  \nonumber \\
\left( 1-v\xi \right) \frac{p^{\prime }}{w} &=&\gamma ^{2}\left( \xi
-v\right) v^{\prime },  \label{fluidx2}
\end{eqnarray}
where a prime indicates derivative with respect to $\xi $. The
pressure and energy density are further related by the equation of
state. According to Eq. (\ref{cs}), we have $p^{\prime
}=c_{s}^{2}e^{\prime }$. Using this relation, Eqs. (\ref{fluidx2})
can be combined to obtain the central equation for the velocity
profile \cite{s82,k85,kkt94}. We obtain
\begin{equation}
j\frac{v}{\xi }=\gamma ^{2}\left( 1-v\xi \right)
\left[ \frac{\mu ^{2}}{c_{s}^{2}}-1\right] v^{\prime },  \label{profile}
\end{equation}
where
\begin{equation}
\mu \left( \xi ,v\right) =\frac{\xi -v}{1-\xi v}.
\end{equation}
From Eqs. (\ref{fluidx2}) we also obtain the equation for the
enthalpy profile,
\begin{equation}
\frac{w^{\prime }}{w}=\left( \frac{1}{c_{s}^{2}}+1\right) \mu \gamma
^{2}v^{\prime },  \label{wprof}
\end{equation}
which is readily integrated \cite{kkt94}. For $c_s^2=1/3$ we have
\begin{equation}
\frac{w_{b}}{w_{a}}=\exp \left[ \int_{\xi _{a}}^{\xi _{b}}4\gamma ^{2}\mu
\left( \xi ,v\right) v^{\prime }d\xi \right] .  \label{enth}
\end{equation}
Notice that, in this model, the equations for the velocity and
enthalpy profiles are the same in both phases.

\subsection{Discontinuities}

\subsubsection{The phase transition front}

In the $\xi$ axis, the bubble wall is at $\xi _{w}= v_{w}$. The
enthalpy and other quantities are discontinuous at $\xi_w$, and so
will be the fluid velocity. \emph{In the reference frame of the
wall}, the fluid comes from the high-T phase with a velocity $v_{+}$,
and goes out into the low-T phase with a velocity $v_{-}$. The
incoming and outgoing flow velocities are related by the conservation
of $T^{\mu \nu }$ across the wall \cite{landau},
\begin{eqnarray}
w_{-}v_{-}^{2}\gamma _{-}^{2}+p_{-} &=&w_{+}v_{+}^{2}\gamma _{+}^{2}+p_{+},
\label{landaua} \\
w_{-}v_{-}\gamma _{-}^{2} &=&w_{+}v_{+}\gamma _{+}^{2}.  \label{landaub}
\end{eqnarray}
Equivalently,
\begin{equation}
v_{+}v_{-}=\frac{p_{+}-p_{-}}{e_{+}-e_{-}},\quad \frac{v_{+}}{v_{-}}=
\frac{e_{-}+p_{+}}{e_{+}+p_{-}}.  \label{landau2}
\end{equation}
Notice that these equations do not depend on $j$. This is because the
surface of discontinuity is locally planar. In the reference frame of
the bubble center, the fluid velocities on each side of the wall are
given by  $\tilde{v}_{\pm }=\mu \left( \xi _{w},|v_{\pm }|\right) $.

According to Eq. (\ref{landaub}), $v_{+}$ and $ v_{-} $ have the same
sign. Indeed, the fluid velocities in the system of the wall must be
both negative. Using the bag EOS, Eqs. (\ref{landau2}) can be
combined to obtain a relation between $v_{+}$, $v_{-}$ and the
parameter
\begin{equation}
\alpha _{+}\equiv \frac{\varepsilon}{a_{+}T_{+}^{4}}.
\end{equation}
We can solve, e.g., for $v_{+}$ as a function of $v_-$ and $\alpha_+$
\cite{s82},
\begin{equation}
v_{+}=\frac{\left( \frac{v_{-}}{2}+\frac{1}{6v_{-}}\right) \pm \sqrt{\left(
\frac{v_{-}}{2}+\frac{1}{6v_{-}}\right) ^{2}+\left( 1+\alpha _{+}\right)
\left( \alpha _{+}-1/3\right) }}{1+\alpha _{+}}.  \label{vmavme}
\end{equation}
Two kinds of hydrodynamical processes may occur at the phase
transition front, corresponding to the $+$ and $-$ signs in Eq.
(\ref{vmavme}), namely, a \emph{detonation}, for which the incoming
flow is supersonic ($\left\vert v_{+}\right\vert
>c_{s}$) and faster than the outgoing flow ($\left\vert
v_{-}\right\vert <\left\vert v_{+}\right\vert $), and a
\emph{deflagration}, with $\left\vert v_{+}\right\vert <c_{s}$ and
$\left\vert v_{-}\right\vert
>\left\vert v_{+}\right\vert $. In either case, the incoming  velocity
$\left\vert v_{+}\right\vert $ has an extremum at $\left\vert
v_{-}\right\vert =c_{s}$, namely, a minimum for detonations and a
maximum for deflagrations. A process with $\left\vert
v_{-}\right\vert =c_{s}$ is called a \emph{Jouguet} detonation or
deflagration. In this case we have $ |v_{+}|=v_{J} \left( \alpha
_{+}\right) $, with
\begin{equation}
v_{J}^{\mathrm{det} \atop \mathrm{def}}\left( \alpha _{+}\right) =
\frac{1\pm \sqrt{\alpha _{+}\left( 2+3\alpha _{+}\right) }}{\sqrt{3}\left(
1+\alpha _{+}\right) }.
\end{equation}
Hence, for detonations we have $c_{s}<v_{J}^{\mathrm{\det }}\left(
\alpha _{+}\right) \leq |v_{+}|,$ and for deflagrations $|v_{+}|\leq
v_{J}^{\mathrm{def}}\left( \alpha _{+}\right) <c_{s}$. The
hydrodynamical process is called \emph{weak} if the velocities
$v_{+}$ and $v_{-}$ are either both supersonic, or both subsonic. In
such a case, $\left\vert v_{-}\right\vert $ lies between $\left\vert
v_{+}\right\vert $ and $c_{s}$. Otherwise, one of the two velocities
is subsonic and the other one is supersonic. Such hydrodynamical
process is called \emph{strong}.

\subsubsection{Shock fronts}

Discontinuities in the same phase may also be needed to satisfy the
boundary conditions. Such discontinuities are called \emph{shock
fronts}.  We shall use the index 1 for fluid variables behind the
shock front at $\xi=\xi_{\mathrm{sh}}$, and the index 2 for variables
in front of the shock. The EOS is the same on both sides of the
discontinuity, and Eqs. (\ref{landau2}) trivially give
\begin{equation}
v_{1}v_{2}=\frac{1}{3},\quad \frac{v_{1}}{v_{2}}=
\frac{3T_{2}^{4}+T_{1}^{4}}{3T_{1}^{4}+T_{2}^{4}},  \label{landaushock}
\end{equation}
where $v_{1}$ and $v_{2}$ are the (negative) fluid velocities
\emph{in the shock frame}.

When dealing with discontinuities, entropy considerations can be
useful for discarding possible processes (see, e.g., \cite{gkkm84}).
Consider a portion of the fluid which passes through a discontinuity
surface. Requiring the entropy of the fluid to increase, one obtains
the condition $s_1v_1\gamma_1\geq s_2v_2\gamma_2$. Using $s=w/T$ and
Eq. (\ref{landaub}), this condition becomes
\begin{equation}
T_2/T_1\geq\gamma_1/\gamma_2 .
\label{entropT}
\end{equation}
For the bag EOS, we have $w_2/w_1=(T_2/T_1)^4$, and we may insert the
inequality (\ref{entropT}) back in Eq. (\ref{landaub}) to
obtain\footnote{Similarly, for the phase transition front we obtain
$v_{+}(1-v_+^2)\leq (a_{-}/a_{+})v_-(1-v_-^2)$.}
\begin{equation}
v_{2}(1-v_2^2)\leq v_1(1-v_1^2).
\end{equation}
Using the relation $v_{1}v_{2}=1/3$,  this condition becomes $\left(
v_{1}^{2}-1/3\right) ^{3}\leq 0$, which implies
\begin{equation}
|v_{1}|<\frac{1}{\sqrt{3}}<|v_{2}|.  \label{entropshock}
\end{equation}

In the frame of the bubble center, the fluid velocities  on each side
of the shock front are given by $\tilde{v}_{1,2}=\mu \left( \xi
_{\mathrm{sh}},|v_{1,2}|\right) $.  According to the condition
(\ref{entropshock}), we have
\begin{equation}
\tilde{v}_{1}>\tilde{v}_{2}.  \label{saltoentro}
\end{equation}
Hence, the fluid velocity must have a negative jump. As a
consequence, $\tilde{v}_{1}$ cannot vanish (otherwise, we would have
$\tilde{v}_{2}< 0$, but, in this reference frame, the fluid
velocities are positive or zero). On the other hand, we may have
$\tilde{v}_{2}=0$. In this case, the velocity of the shock front is
given by $\xi _{\mathrm{sh}}=-v_{2}$, and the first of Eqs.
(\ref{landaushock}) gives
\begin{equation}
\tilde{v}_{1}=\frac{3\xi _{\mathrm{sh}}^{2}-1}{2\xi _{\mathrm{sh}}}.
\label{v1t}
\end{equation}
Equivalently,
\begin{equation}
\xi _{\mathrm{sh}}=\frac{\tilde{v}_{1}}{3}+\sqrt{\left(
\frac{\tilde{v}_{1}}{3}\right) ^{2}+\frac{1}{3}},  \label{xish}
\end{equation}
which implies that the shock is supersonic.

\subsection{Kinds of solutions}

Equation (\ref{profile}) can be solved numerically for the spherical
and cylindrical cases, and analytically for the planar case. General
solution curves for the three symmetries can be found in  Ref.
\cite{k85} (for the planar case, see section \ref{planar} below). The
velocity profile of the fluid must  fulfil the discontinuity
conditions at the bubble wall, and is constructed by matching
different solutions of Eq. (\ref{profile}). The boundary conditions
are that the fluid is at rest far ahead of the phase transition
surface (where no information of the phase transition has arrived
yet) and far behind that surface (near the center of the bubble). Far
in front of the phase transition front, the temperature is still
$T_N$. Therefore, the boundary condition for the enthalpy density is
that it takes the value
\begin{equation}
w_{N}=\frac{4}{3}a_{+}T_{N}^{4}  \label{wn}
\end{equation}
far in front of the bubble wall.

Not all of the aforementioned hydrodynamical processes will be
realized in a phase transition. It is known that strong detonations
are not possible, since they cannot satisfy the boundary conditions
\cite{landau,s82,l94}. It has been argued in Ref. \cite{s82} that
weak detonations are also impossible, like in the case of chemical
burning \cite{landau}. As a consequence, we would only have Jouguet
detonations, and the velocity $v_{w}=v_{J}$ would be completely
determined by hydrodynamics and would not depend on microphysics.
However, it has been shown \cite{l94} that this is not true in the
case of phase transitions, where the situation is similar to that of
condensation discontinuities \cite{landau} rather than chemical
burning. Consequently, only strong detonations are forbidden.

Strong deflagrations do not seem to be realized either. In Ref.
\cite{l94} it was argued that, like in the case of chemical burning,
they are forbidden by entropy considerations \cite{landau}. However,
in Ref. \cite{kl95}, it was shown that this proof is not valid for
cosmological phase transitions. Nevertheless, a mechanical
instability argument against strong deflagrations \cite{landau} seems
to be valid also for cosmological phase transitions \cite{kl95}.
Numerical calculations \cite{ikkl94,kl95} support this assertion. On
the other hand, in Ref.  \cite{kl95} supersonic Jouguet deflagrations
were shown to exist.

As a consequence, three kinds of solutions seem to be realized in
nature, namely, weak detonations, subsonic weak deflagrations, and
supersonic Jouguet deflagrations. In section \ref{planar}, for the
planar case, we argue that these are the only possible solutions (the
argument is simpler for the planar case, but can be straightforwardly
generalized to the other symmetries). In the rest of the present
section we describe these solutions and calculate their profiles and
the efficiency factor (the spherical case has been studied recently
in Ref. \cite{ekns10}).

\subsubsection{Detonations}

\begin{figure}[hbt]
\epsfysize=5.5cm \leavevmode \epsfbox{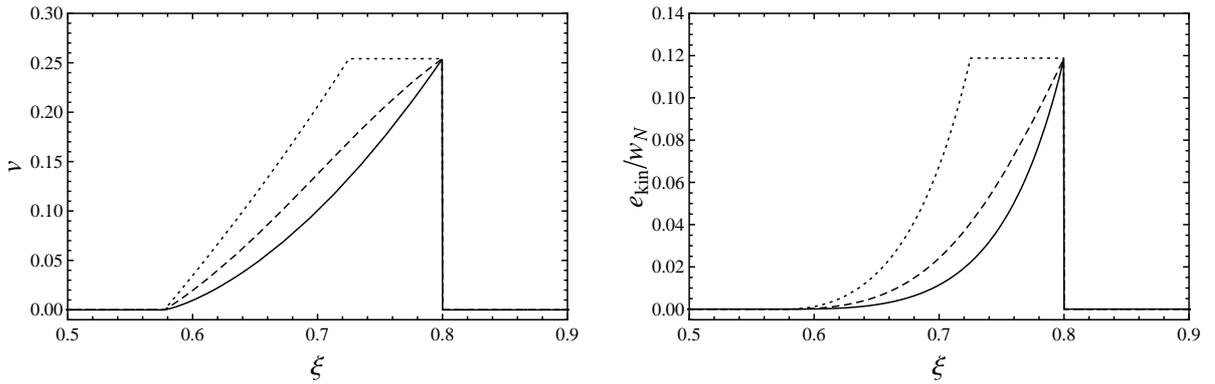}
\caption{Left: the fluid velocity profile of a detonation with $v_{w}=0.8$ and $\alpha
_{N}=0.1$ for a spherical wall (solid), a cylindrical wall (dashed), and a
planar wall (dotted). Right: the corresponding kinetic energy density
profiles.}
\label{profsdeto}
\end{figure}

For detonations we have $|v_+|\geq v_J^{\mathrm{det}}$, and the wall
moves supersonically with respect to the fluid in front of it. Thus,
outside a detonation bubble the fluid has not yet been perturbed.
Hence the fluid velocity vanishes and the temperature is still that
at which the bubble nucleated, i.e., $\tilde{v}_{+}=0$ and
$T_{+}=T_{N}$. Therefore, we have
\begin{equation}
v_{+}=-\xi_w , \;\;\alpha _{+}=\alpha _{N},
\end{equation}
where
\begin{equation}
\alpha _{N}\equiv\frac{\varepsilon}{a_{+}T_{N}^{4}} .
\end{equation}
As a consequence, we have $\xi_w\geq v_J^{\mathrm{det}}$. The
velocity profile of a detonation is shown in Fig. \ref{profsdeto}
(left) for the three wall geometries. The bubble wall is followed by
a rarefaction wave which ends at $\xi =c_{s}$. The velocity profile
is determined by the boundary condition $v\left( \xi _{w}\right)
=\tilde{v}_{-}$, with
\begin{equation}
\tilde{v}_{-}=\mu \left( \xi _{w},|v_{-}|\right) .  \label{vmet}
\end{equation}
The velocity $v_{-}$ is given by the inverse of relation
(\ref{vmavme}),
\begin{equation}
\left\vert v_{-}\right\vert =\left( \frac{\left\vert v_{+}\right\vert \left(
1+\alpha _{+}\right) }{2}+\frac{\frac{1}{3}-\alpha _{+}}{2\left\vert
v_{+}\right\vert }\right) \pm \sqrt{\left( \frac{\left\vert v_{+}\right\vert
\left( 1+\alpha _{+}\right) }{2}+\frac{\frac{1}{3}-\alpha _{+}}{2\left\vert
v_{+}\right\vert }\right) ^{2}-\frac{1}{3}}.  \label{vmevma}
\end{equation}
Weak detonations (i.e., $|v_{-}|>c_{s}$) correspond to the $+$ sign
in Eq (\ref{vmevma}) and strong detonations ($|v_{-}|<c_{s}$)
correspond to the $-$ sign\footnote{In Eq. (\ref{vmevma}), $|v_{-}|$
is real only for $|v_{+}|\geq v_J^{\mathrm{det}}$ (corresponding to
detonations) or $|v_{+}|\leq v_J^{\mathrm{def}}$ (corresponding to
deflagrations).}. The two branches match at the Jouguet point
$|v_{+}|= v_J^{\mathrm{det}}, |v_{-}|=c_{s}$. As we mentioned, the
process must be a weak (or, at most, Jouguet) detonation for
compatibility with the boundary conditions.

The enthalpy profile is given by Eq. (\ref{enth}) with the condition
$w\left( \xi _{w}\right) =w_{-}$. The value $w_{-}$ just behind the
wall is related to $ w_{+}=w_{N}$ through Eq. (\ref{landaub}),
\begin{equation}
w_{-}=\frac{v_{w}\gamma _{w}^{2}}{|v_{-}|\gamma _{-}^{2}}w_{N}.  \label{wme}
\end{equation}
The efficiency factor is obtained by integrating the kinetic energy
density $ e_{kin}=wv^{2}\gamma ^{2}$. Figure \ref{profsdeto} (right)
shows the kinetic energy density profile for the three wall
geometries.

\subsubsection{Subsonic deflagrations}

For $\xi _{w}<c_{s}$, the wall is preceded by a shock front at $\xi
_{\mathrm{sh}}>c_{s}$.  Behind the wall the fluid velocity vanishes.
Hence, we have $\tilde{v}_{-}=0$ and (since $\tilde{v}_+>0$)
$|v_{+}|<|v_{-}|=\xi_w<c_{s}$, i.e., the hydrodynamical process is a
weak deflagration (in the limiting case $\xi _{w}=c_{s}$, we have a
Jouguet deflagration).  The fluid velocity vanishes also beyond the
shock front. Thus, we have $\tilde{v}_{2}=0$ and $\alpha_2=\alpha_N$.

\begin{figure}[hbt]
\centering \epsfysize=5.5cm \leavevmode \epsfbox{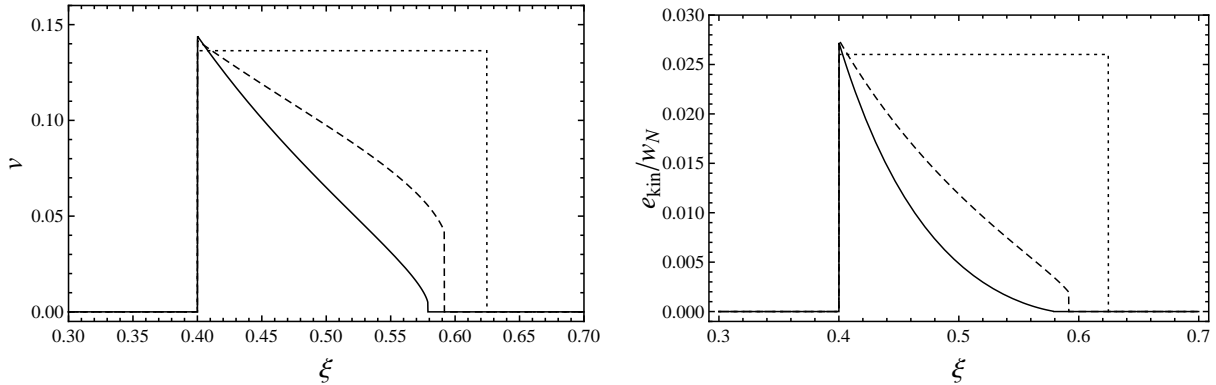}
\caption{Left: the fluid velocity profile of a  deflagration with
$v_{w}=0.4$
and $\alpha _{N}=0.1$ for a spherical wall (solid), a cylindrical wall
(dashed), and a planar wall (dotted). Right: the corresponding kinetic
energy density profiles.}
\label{profsdefla}
\end{figure}

The velocity profile of the shock wave between $\xi_w$ and
$\xi_{\mathrm{sh}}$ is given by a solution of Eq. (\ref{profile})
(see Fig. \ref{profsdefla}). One can choose as boundary condition the
value of the velocity $\tilde{v}_{+}$ in front of the wall, or the
value $\tilde{v}_{1}$ behind the shock. The velocity $\tilde{v}_{+}$
is given by $\mu \left( \xi _{w},|v_{+}|\right) $, where $v_{+}$ is
given by Eq. (\ref{vmavme}) as a function of $\alpha _{+}$ and of $
v_{-}=-v_{w}$. However, in the present case  $\alpha _{+}$ does not
equal $\alpha _{N}$, since the shock wave reheats the fluid in front
of the wall. The temperatures at each end of the shock wave are
related by
\begin{equation}
\alpha _{+}=\frac{w_{1}}{w_{+}}\alpha _{1},  \label{alfamalfa1}
\end{equation}
with $w_{1}/w_{+}$ given by Eq. (\ref{enth}),
\begin{equation}
\frac{w_{1}}{w_{+}}=\exp \left[ \int_{\tilde{v}_{+}}^{\tilde{v}_{1}}4\gamma
^{2}\mu \left( \xi ,v\right) dv\right] .  \label{wmaw1}
\end{equation}
The fluid velocities  $\tilde{v}_{+}$ and $\tilde{v}_{1}$ are related
by the fluid equation (\ref{profile}). Let us denote  $v\left( \xi
;\tilde{v}_{1}\right) $ the solution  with boundary condition
$v\left( \xi _{\mathrm{sh}}\right) =\tilde{v}_{1}$, with $\xi
_{\mathrm{sh}}$ depending on $\tilde{v}_{1}$ through Eq.
(\ref{xish}). Then, evaluating at $\xi =\xi _{w}$, we obtain
$\tilde{v}_{+}$ as a function of $\tilde{v}_{1}$ and $\xi _{w}$,
\begin{equation}
\tilde{v}_{+}=v\left( \xi _{w};\tilde{v}_{1}\right) . \label{vmat}
\end{equation}
Equations (\ref{alfamalfa1}-\ref{vmat}) give $\alpha _{+}$ as a
function of $\alpha _{1}$, $\xi _{w}$ and $\tilde{v}_{1}$. The
variables $ \alpha _{1}$ and $\alpha _{2}=\alpha _{N}$ on each side
of  the shock discontinuity are related by Eqs. (\ref{landaushock})
with $v_2=-\xi_\mathrm{sh}$. We obtain
\begin{equation}
\alpha _{1}=
\frac{3\left( 1-\xi _{\mathrm{sh}}^{2}\right)}{9\xi _{\mathrm{sh}}^{2}-1}
\alpha _{N}.  \label{alfa1xish}
\end{equation}
Equivalently, using Eq. (\ref{xish}), we have
\begin{equation}
\alpha _{1}=\frac{\tilde{\gamma}_{1}^{2}}{3}\left( 3+5\tilde{v}_{1}^{2}-4
\tilde{v}_{1}\sqrt{3+\tilde{v}_{1}^{2}}\right) \alpha _{N}.  \label{alfa1}
\end{equation}
Thus, we have $ \alpha _{+}$ as a function of $\alpha _{N}$, $\xi
_{w}$, and either $\tilde{v}_{1}$ or $\xi _{\mathrm{sh}}$. On the
other hand, the variables at the wall discontinuity are related by
Eq. (\ref{vmavme}) or Eq. (\ref{vmevma}) or, equivalently, by
\begin{equation}
\alpha _{+}=\gamma _{+}^{2}\left( v_{+}^{2}+\frac{1}{3}-v_{+}v_{-}-\frac{1}{3}
v_{+}/v_{-}\right) .  \label{alfamadefla0}
\end{equation}
In the present case, we have $ v_{-}=-\xi _{w}$, $v_{+}=\left(
\tilde{v}_{+}-\xi _{w}\right) /\left( 1- \tilde{v}_{+}\xi _{w}\right)
$, and we obtain
\begin{equation}
\alpha _{+}=\frac{\tilde{\gamma}_{+}^{2}\tilde{v}_{+}}{3v_{w}}\left( 2v_{w}
\tilde{v}_{+}+1-3v_{w}^{2}\right) .  \label{alfamadefla}
\end{equation}
Inserting Eq. (\ref{alfamadefla}) in  Eq. (\ref{alfamalfa1}),  we
eliminate $\alpha_+$ and we can solve for $\tilde{v}_{1}$ as a
function of $\xi _{w}$ and $\alpha _{N}$. Searching for
$\tilde{v}_{1}$ numerically, implies the evaluation of all the
quantities which appear in Eqs. (\ref{alfamalfa1}-\ref{alfamadefla}),
for several values of $\tilde{v}_{1}$. Such evaluation involves
numerically solving the differential equation which gives
$\tilde{v}_{+}$ as a function of $\tilde{v}_{1}$ and then performing
numerically the integral in Eq. (\ref{wmaw1}). As we shall see in
section \ref{planar}, in the planar case Eqs.
(\ref{alfamalfa1}-\ref{alfamadefla}) reduce to a single, algebraic
equation, which can be solved analytically.

Once the value of $\tilde{v}_{1}$ is found, one can compute the
velocity and enthalpy profiles and perform the integral of the
kinetic energy density to obtain the efficiency factor. The enthalpy
profile is given by Eq. (\ref{enth}) and determined by the condition
$w\left( \xi _{\mathrm{sh}}\right) =w_{1}$, where $w_{1}=\left(
\alpha _{N}/\alpha _{1}\right) w_{N}$, with $\alpha _{N}/\alpha _{1}$
given by Eq. (\ref{alfa1xish}). Thus, we have
\begin{equation}
w_{1}=
\frac{9\xi _{\mathrm{sh}}^{2}-1}{3\left( 1-\xi _{\mathrm{sh}}^{2}\right) }
w_{N}.
\label{w1}
\end{equation}
Figure \ref{profsdefla} shows the velocity and kinetic energy density
profiles for a subsonic deflagration.

\subsubsection{Supersonic deflagrations}

\begin{figure}[hbt]
\centering \epsfysize=5.5cm \leavevmode \epsfbox{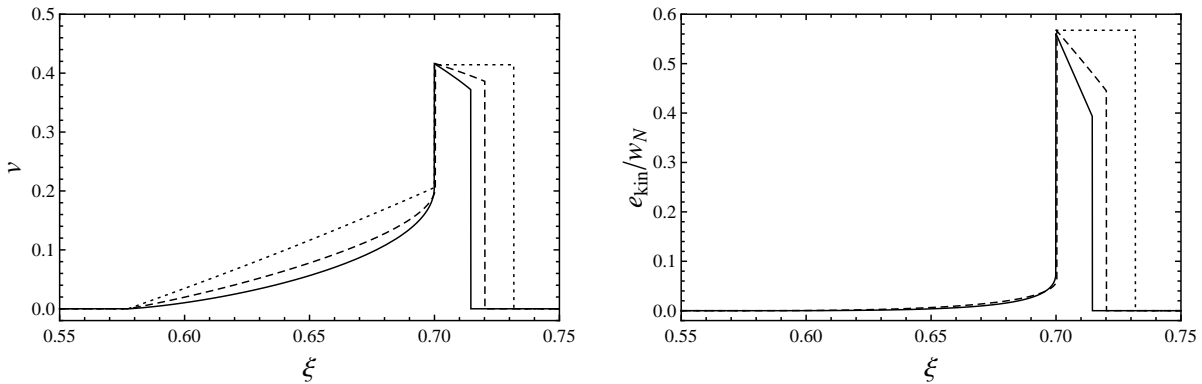}
\caption{Left: the fluid velocity profile of a  deflagration with
$v_{w}=0.7 $ and $\alpha _{N}=0.1$ for a spherical wall (solid), a
cylindrical wall (dashed), and a planar wall (dotted). Right: the
corresponding kinetic energy density profiles.
}
\label{profshibri}
\end{figure}

For $c_{s}<\xi _{w}<v_{J}^{\det }$, the phase transition front is
preceded by a shock front at $\xi_{\mathrm{sh}}>\xi_w$, and is
followed by a rarefaction wave solution which vanishes at $\xi=c_s$.
Therefore, both velocities $\tilde{v}_{+}$ and $\tilde{v}_{-}$ are
non-vanishing (see Fig. \ref{profshibri}). The hydrodynamic process
is a Jouguet deflagration, i.e., $|v_{+}|<|v_{-}|=c_{s}$ \cite{kl95}.
The wall velocity $\xi _{w}=\left( \tilde{v}_{-}+c_{s}\right) /\left(
1+ \tilde{v}_{-}c_{s}\right) $ is supersonic and depends on the value
of $ \tilde{v}_{-}$. For $ \tilde{v}_{-}=0$ this solution matches the
``ordinary'' deflagration considered before. As $\xi_w$ approaches
$v_{J}^{\det }$, the shock wave gets shorter, i.e.,
$\xi_{\mathrm{sh}}\to \xi _{w}$, and the profile matches that of the
detonation considered before.

The calculation of the boundary value $\tilde{v}_{1}$ for the shock
wave profile is very similar to that of the ordinary deflagration.
Indeed, Eqs. (\ref{alfamalfa1}-\ref{alfamadefla0}) hold in this case.
Since the deflagration is now Jouguet, we have the condition
$v_{-}=-c_{s}$ instead of $v_{-}=-v_{w}$. As a consequence, Eq.
(\ref{alfamadefla0}) becomes $\alpha _{+}=\gamma _{+}^{2}\left(
v_{+}+1/\sqrt{3}\right) ^{2}$, and Eq. (\ref{alfamadefla}) gets
replaced by
\begin{equation}
\alpha _{+}=\frac{\tilde{\gamma}_{+}^{2}\tilde{\gamma}_{w}^{2}}{3}\left( 1-
\sqrt{3}v_{w}-\tilde{v}_{+}(v_{w}-\sqrt{3})\right) ^{2}.  \label{alfamahibri}
\end{equation}
From Eqs. (\ref{alfamalfa1}) and (\ref{alfamahibri}) we eliminate
$\alpha_+$  and we obtain $\tilde{v}_{1}$ as a function of $\xi_w$
and $\alpha_N$ as before. The enthalpy profile of the shock wave is
determined by the value $w_1$ at the shock, given by Eq. (\ref{w1}).

The profile of the rarefaction wave is similar to that of the
detonation, with the boundary condition $v\left( \xi _{w}\right)
=\tilde{v}_{-}$, with $ \tilde{v}_{-}$ now given by
\begin{equation}
\tilde{v}_{-}=\frac{\xi _{w}-c_{s}}{1-\xi _{w}c_{s}}.  \label{vmethib}
\end{equation}
The enthalpy profile behind the wall is determined by the condition
$w\left( \xi _{w}\right) =w_{-}$. The value of the enthalpy just
behind the wall is now given by
\begin{equation}
w_{-}=(2/\sqrt{3})|v_{+}|\gamma _{+}^{2}w_{+},  \label{wmehib}
\end{equation}
with  $v_{+}$ given by
\begin{equation}
v_{+}=\left( \tilde{v}_{+}-\xi _{w}\right) /\left( 1-\tilde{v}_{+}\xi
_{w}\right),
\end{equation}
and  $\tilde{v}_{+}$ and  $w_{+}$ are obtained from the shock wave
profile. Figure \ref{profshibri} shows the profile of a supersonic
deflagration for the three geometries.

\subsection{Energy injected into the plasma}

The latent heat released at the phase transition fronts spreads out
in the plasma. Part of this energy causes reheating, and another part
causes bulk motions. Some of the consequences of the phase transition
will depend on the thickness of the plasma shell where the energy is
concentrated. The issue of reheating was addressed, e.g., in Refs.
\cite{h95,m01} for the electroweak phase transition. The size of the
regions of reheated plasma affects the dynamics of the phase
transition and, as a consequence, the baryogenesis mechanism. Here we
shall focus on the energy in bulk motions of the plasma.

\begin{figure}[hbt]
\centering \epsfxsize=10cm \leavevmode \epsfbox{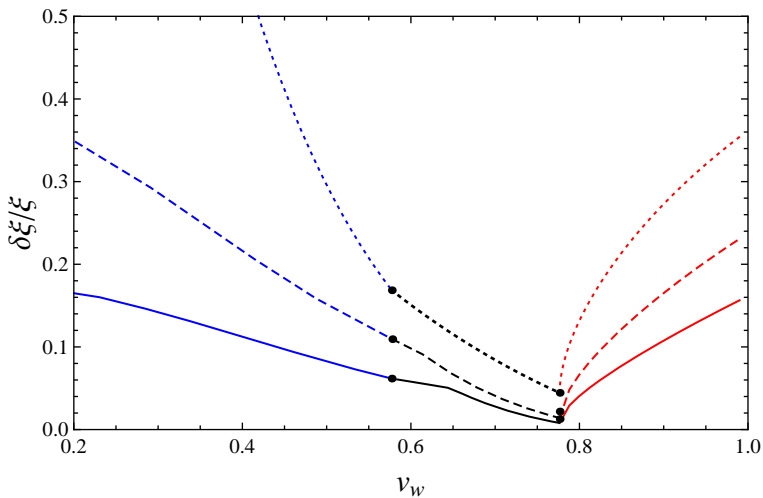}
\caption{The thickness $\delta \xi$ of the region around the wall
inside which the kinetic energy
density decreases to a half of its maximum value, for
$\alpha _{N}=0.1$ as a function of $v_{w}$. Solid lines correspond to
spherical bubbles, dashed lines correspond to cylindrical bubbles, and
dotted lines correspond to planar walls. Subsonic deflagrations are plotted
in blue, supersonic deflagrations are in black, and detonations in red.}
\label{figdx}
\end{figure}

The bubble walls set the fluid moving forward with velocities
$\tilde{v}_+$ and $\tilde{v}_-$. In the left panels of Figs.
\ref{profsdeto}, \ref{profsdefla}, and \ref{profshibri} we observe
that these velocities are very similar for different wall geometries.
However, away from the wall the fluid velocity decays faster for more
symmetric bubbles, since in that case there is more room for the
energy to get distributed. For illustrative purposes, we define the
shell where kinetic energy is concentrated as the region around the
wall where the kinetic energy density remains higher than half the
maximum on each side of the wall. Thus, the thickness is given by
$\delta \xi=\xi_+-\xi_-$, where $\xi_+$ and $\xi_-$ are determined by
the condition $e_{\mathrm{kin}}(\xi_{\pm})= 0.5\,
w_{\pm}\tilde{v}_{\pm}^2\tilde{\gamma}_{\pm}^2$. We plot the value of
$\delta \xi$ in Fig. \ref{figdx}. The shape of the curves is
different from those given in Ref. \cite{ekns10} for the spherical
case, since we have defined $\delta \xi$ differently. Nevertheless,
the general structure is similar: the thickness is larger for
subsonic deflagrations than for detonations, and is quite small for
Jouguet solutions. We also see that the energy is more widely
distributed in the planar case, especially for weak deflagrations.

Some of the cosmological remnants of the phase transition do not
depend on the thickness of the region of perturbed fluid around the
wall. This is the case, e.g., when the generating mechanism is based
on the turbulence produced by the colliding bubble walls. Eddies are
formed at all size scales up to the size of the largest bubbles.

\begin{figure}[hbt]
\centering \epsfxsize=12cm \leavevmode \epsfbox{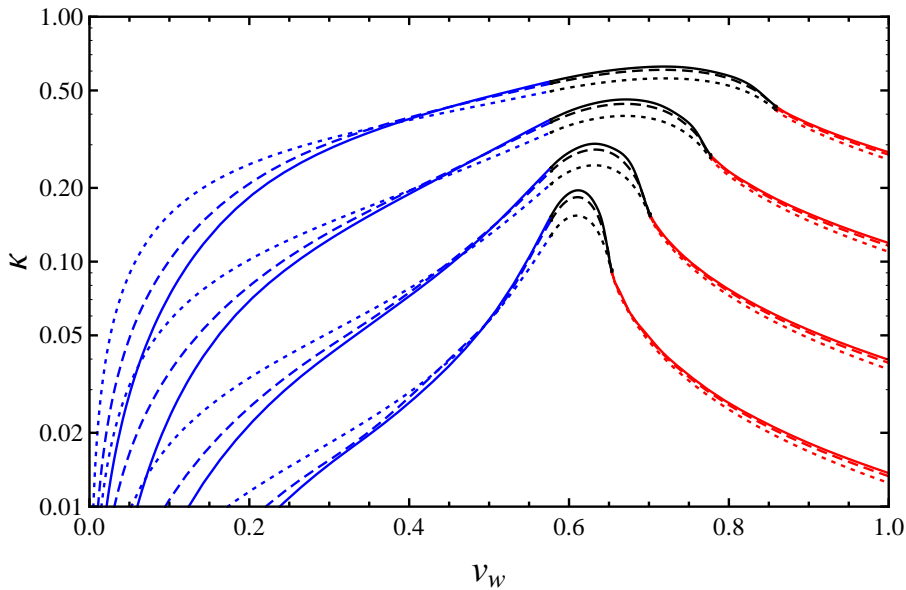}
\caption{The efficiency factor $\kappa $
as a function of $v_{w}$ for $\alpha _{N}=0.01,0.03,0.1,0.3$.
Solid lines correspond to the spherical case,
dashed lines correspond to the cylindrical case, and dotted lines correspond
to the planar case. Subsonic deflagrations are plotted in blue,
supersonic deflagrations are in black, and detonations in red.}
\label{figkappa}
\end{figure}

It is usually assumed that detonation walls cause a stronger
disturbance of the fluid than deflagration walls, since detonations
have higher velocities. For weak detonations, however, as pointed out
in Ref. \cite{ms10}, the higher the wall velocity, the smaller the
disturbance of the fluid. Thus, the strongest disturbance is caused
by the Jouguet detonation (which is the case usually considered in
the GW literature). However, this is just a limiting case; real
detonations are generally weak. Furthermore, as noticed in Ref.
\cite{m08}, ordinary deflagrations may cause important perturbations
in the fluid if they are close to the Jouguet limit $v_{w}=c_{s}$. In
Fig. \ref{figkappa} we give the efficiency factor $\kappa $ as a
function of $v_{w}$ for several values of $\alpha_{N}$. We calculated
$\kappa$ numerically for the spherical and cylindrical cases, and
analytically for the planar case (see section \ref{planar}). For the
spherical case (solid lines), our results agree with the numerical
fits provided in Ref. \cite{ekns10}.  As expected, for fixed $\alpha
_{N}$, the efficiency is larger for stronger solutions (i.e.,
solutions which are closer to Jouguet processes). Thus, for subsonic
deflagrations the efficiency factor increases with the wall velocity,
whereas for detonations $\kappa $ decreases with $v_{w}$. The
efficiency peaks for supersonic deflagrations, which are Jouguet
processes. It is interesting to notice that even subsonic
deflagrations can give larger efficiency factors than detonations.

We see that the different geometries give in general similar values
of $\kappa $, except for the case of small wall velocities
($v_{w}\lesssim 0.2$). This case will not be interesting in general
for GW generation, since the efficiency factor is small. For faster
walls, the three curves are quite close; they only separate somewhat
for supersonic deflagrations (in Fig. \ref{figkappa}, the differences
at the top of the curves are less than a 20\%). In any case, it is
clear from Fig. \ref{figkappa} that the difference between geometries
will always be at most an $\mathcal{O}\left( 1\right) $ factor,
except for uninterestingly small values of $\kappa $.

\section{Planar walls: analytic results \label{planar}}

\begin{figure}[hbt] \centering \epsfysize=6cm \leavevmode
\epsfbox{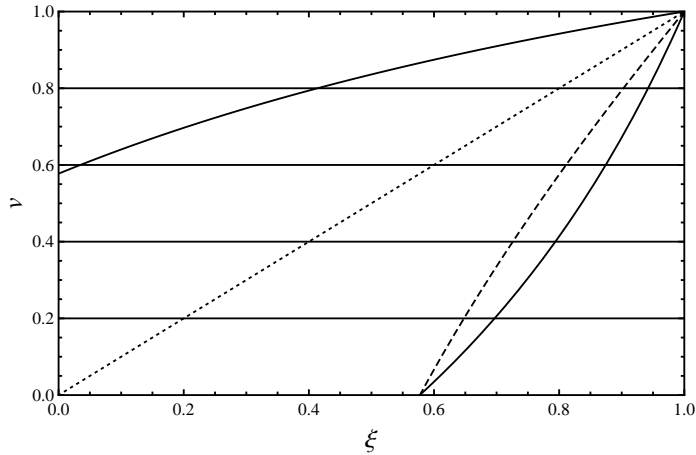} \caption{Solutions to the fluid velocity equation
for the planar case (solid lines). The dotted line is the
curve $v=\xi$. Physical solutions are below this curve. The dashed
line corresponds to the value of
the fluid velocity $\tilde{v}_{1} $ at the shock front.}
\label{figfluid}
\end{figure}

In this section we calculate analytically the kinetic energy density
and the efficiency coefficient for planar walls. In this case the
``bubble'' consists of two planar walls  at positions $x=v_w t$ and
$x=-v_w t$, and is equivalent to a bubble in 1+1 dimensions. The
system is symmetric under reflection through a plane, and we need
only consider the wall moving to the right. The absence of a length
scale in the problem implies that the profiles depend only on
$\xi=x/t$. The solutions of the fluid velocity equation for the
planar case are well known. For $j=0$, Eq. (\ref{profile}) implies
either that $v^{\prime }\left( \xi \right) \equiv 0$ or that $\mu
\left( \xi ,v\right) \equiv \pm c_{s}$. The latter implies that
$v=\mu \left( \xi ,\pm c_{s}\right)$. These solutions are shown in
Fig. \ref{figfluid} for $\xi\geq 0$ and $v\geq 0$. We need only
consider that quadrant since the wall at $\xi_w=v_w$ sets the fluid
moving forward (the reflected profiles around the opposite
wall\footnote{More complicated profiles will arise if one considers
two bubbles nucleated at a distance $d$ \cite{kk86}. In particular,
this separation introduces a new length scale in the problem.} at
$\xi=-v_w$ are constructed with the solutions for $\xi<0$ and $v<0$).
As we shall see, the solution $v=\mu \left( \xi ,- c_{s}\right)$ will
not take part in the fluid profile. Indeed, the possible values of
the fluid velocity will be those below the curve $v=\xi$ (in dots in
Fig. \ref{figfluid}) due to the fact that, at the wall, the fluid
velocity fulfils $\tilde{v}_{\pm}<\xi_w$ (since in the wall frame we
have $v_{\pm}<0$). As a consequence, the physical solutions are
either the constants or the ``rarefaction''
\begin{equation}
v_{\mathrm{rar}}\left( \xi \right) =\frac{\xi -c_{s}}{1-c_{s}\xi } .
\label{vrar}
\end{equation}
The solution $v_{\mathrm{rar}}$ is positive only for $\xi \geq
c_{s}$. Between $\xi=c_s$ and $\xi=1$, $v_{\mathrm{rar}}(\xi)$ grows
monotonically from $v=0$ to $v=1$.

The enthalpy is readily obtained as well. For the case $v^{\prime
}\equiv 0$ the enthalpy is a constant. For the case $\mu \equiv
c_{s}$, the integral in Eq. (\ref{enth}) is simple; given the
condition $w(\xi_a)=w_a$, we obtain
\begin{equation}
\frac{w}{w_{a}}=\left( \frac{1-{v}_{a}}{1+{v}_{a}}\frac{1+v}{1-v}\right)
^{2/\sqrt{3}}. \label{enthpl}
\end{equation}

The velocity profile must be constructed by matching  solutions  in
such a way that the discontinuity conditions at the bubble wall and
the boundary conditions are satisfied (shock discontinuities may also
be needed). Before obtaining the analytical results, we shall examine
all the possible profiles and argue that only the three kinds of
solutions considered in section \ref{fluid} are acceptable. The
argument is simpler for the planar case, but the generalization to
the other wall geometries is straightforward.

\subsection{Hydrodynamic processes and fluid profiles}

The fluid velocity must vanish far in front of the wall, where no
signal from the phase transition front has arrived yet. Since there
are no decreasing solutions, we see that, no matter what the position
$\xi_w$ of the wall is, the fluid velocity must have a jump from
$v>0$ to $v=0$. This discontinuity can be either at $\xi= \xi_w$ or
at some point $\xi _{\mathrm{sh}}$ between $ \xi _{w}$ and $1$.

In the former case, we have $v=0$ for all $\xi
>\xi _{w}$. In particular, $\tilde{v}_{+}=0$
and, therefore, $\xi _{w}=|v_{+}|$. Besides, the fluid velocity
behind the wall must be nonvanishing, i.e., $\tilde{v}_{-}>0$. Thus,
we have $\tilde{v}_{-}> \tilde{v}_{+}$ and, hence, $|v_{+}|>|v_{-}|$.
Therefore, the hydrodynamical process is a \textbf{detonation}. As a
consequence, the wall is {supersonic}, since $\xi_w=|v_{+}|\geq
v_{J}^{\det }>c_{s}$. The fluid velocity must vanish also (by
symmetry) at the bubble center. According to Eq. (\ref{saltoentro}),
the velocity cannot have positive jumps in the same phase. Therefore,
$v\left( \xi \right) $ must grow continuously from $0$ to
$\tilde{v}_{-}$. The only possibility is that  $v\equiv 0$ for $\xi
<c_{s}$ and $v= v_{\mathrm{rar}}\left( \xi \right) $ for $\xi \geq
c_{s}$. At some point between $c_{s}$ and $\xi _{w}$, the solution
$v_{\mathrm{rar}}\left( \xi \right) $ may be matched again to a
constant $v\equiv \tilde{v}_{-}$, or it may continue growing until
$\xi =\xi _{w}$. In any case, we have $\tilde{v}_{-}\leq
v_{\mathrm{rar}}\left( \xi _{w}\right) =\mu \left( \xi
_{w},c_{s}\right) $, which implies that $|v_{-}|\geq c_{s}$. Hence,
\emph{the detonation can only be weak or Jouguet}.

If the fluid velocity does not vanish in front of the wall (i.e.,
$\tilde{v} _{+}>0$), then it is clear from Fig. \ref{figfluid} that
the profile must have a shock discontinuity at some point $\xi
_{\mathrm{sh}}>\xi _{w}$, so that $v\equiv 0$ for
$\xi>\xi_{\mathrm{sh}}$. The fluid velocity $\tilde{v}_{1}$ at the
shock front is given by Eq. (\ref{v1t}) and shown in a dashed line in
Fig. \ref{figfluid}. We see that the solution $v_\mathrm{rar}$ lays
completely on the right of the curve of
$\tilde{v}_1(\xi_{\mathrm{sh}})$. Therefore, between  $\xi _{w}$ and
$\xi _{\mathrm{sh}}$  the solution must be a constant,  $ v\equiv
\tilde{v}_{+}=\tilde{v}_{1}$.

Behind the wall, $v$ can grow continuously from $0$ only if the wall
is supersonic.

If the wall is {subsonic}, then the fluid velocity must be $v\equiv
0$ for $\xi <\xi _{w}$ (otherwise we would need a positive jump). In
this case we have $\tilde{v}_{-}<\tilde{v}_{+}$, which implies
$|v_{+}|<|v_{-}|$, and the process is a \textbf{deflagration}.
Furthermore, since $\tilde{v}_{-}=0$ we have $|v_{-}|=\xi _{w}\leq
c_{s}$, i.e., the deflagration is \emph{weak} or, at most,
\emph{Jouguet}.

If the wall is {supersonic}, we still have solutions for which
$v\equiv 0$ behind the wall. In this case, the condition
$\tilde{v}_{-}=0$ implies a strong deflagration ($|v_{-}|=\xi
_{w}>c_{s}$). However, numerical simulations indicate that strong
deflagrations are unstable \cite{ikkl94,kl95}. Notice that, since
$\xi _{w}$ is now $>c_{s}$, we may have a non-vanishing fluid
velocity $\tilde{v}_{-}$ behind the wall, as in the case in which
$\tilde{v}_{+}=0$. Like in that case, we have the condition
$\tilde{v}_{-}\leq v_{\mathrm{rar}}\left( \xi _{w}\right) $ but,
instead of $\tilde{v}_{+}=0$, we now have
$\tilde{v}_{+}=\tilde{v}_{1}\left( \xi _{\mathrm{sh}}\right)
>v_{\mathrm{rar}}\left( \xi _{w}\right)
$ (see Fig. \ref{figfluid}). Hence,  we have $\tilde{v}_{-}<
\tilde{v}_{+}$ and, thus, $|v_{+}|<|v_{-}|$. Therefore, the present
case is again a \textbf{deflagration}, not a detonation. Now
$\tilde{v}_{-}$ can take any value, with the only condition
$\tilde{v}_{-}\leq v_{\mathrm{rar}}\left( \xi _{w}\right) $, which
implies that $|v_{-}|\geq c_{s}$, i.e., the deflagration must be
strong or {Jouguet}. Of all these solutions, though, one expects that
the stable one will be that which causes the smallest perturbation of
the fluid, i.e., the \emph{Jouguet deflagration} $|v_{-}|=c_{s}$.
This is supported by numerical simulations. As we shall see, the
supersonic deflagration matches the detonation solution at
$\xi_w=v_{J}^{\det }$.

Thus, a subsonic phase transition front always propagates as a weak
deflagration and is preceded by a shock wave. A supersonic phase
transition front is always followed by a rarefaction wave, and may
propagate either as a Jouguet deflagration preceded by a shock front,
or as a detonation, without a shock wave.

\subsection{Detonations}

For detonations, the fluid velocity is given by
$v=v_{\mathrm{rar}}\left( \xi \right) $ between $\xi =c_{s}$ and a
certain $\xi _{0}\leq\xi_w$, and by $v\equiv\tilde{v}_{-}$ between
$\xi _{0}$ and $\xi _{w}$ (see Fig. \ref{profsdeto}).  The matching
condition $v_{\mathrm{rar}}\left( \xi _{0}\right) = \tilde{v}_{-}$
determines the value of $\xi _{0}$ as a function of $\tilde{v}_{-}$,
\begin{equation}
\xi _{0}=\frac{\tilde{v}_{-}+c_{s}}{1+\tilde{v}_{-}c_{s}}.
\end{equation}
The velocity $\tilde{v}_{-}$ is given by Eqs. (\ref{vmet}) and
(\ref{vmevma}) as a function of $\alpha _{+}=\alpha _{N}$ and
$v_{+}=-\xi _{w}$. The enthalpy is a constant $w\equiv w_{-}$ for
$\xi _{0}<\xi <\xi _{w}$, where $w_{-}$ is given by Eq. (\ref{wme})
as a function of $w_{N}$ and $\xi _{w}$.  Between $c_{s}$ and $\xi
_{0}$, the enthalpy is given by Eq. (\ref{enthpl}), with the
condition $w(\xi_0)=w_-$. Inserting the velocity profile
(\ref{vrar}), we obtain
\begin{equation}
w=w_{-}\left( \frac{1-c_{s}}{1+c_{s}}\frac{1-\tilde{v}_{-}}{1+\tilde{v}_{-}}
\frac{1+\xi }{1-\xi }\right) ^{2/\sqrt{3}}.  \label{entrar}
\end{equation}
Using Eqs. (\ref{entrar}) and (\ref{vrar}), the efficiency
coefficient (\ref{kappa}) is given by
\begin{equation}
\kappa =\frac{w_{-}}{\xi _{w}\varepsilon }\left[ \frac{\tilde{v}_{-}^{2}
\left( \xi _{w}-\xi _{0}\right) }{1-\tilde{v}_{-}^{2}}+\frac{3}{2}
\left( 2-\sqrt{3}\right) ^{2/\sqrt{3}}\left(
\frac{1-\tilde{v}_{-}}{1+\tilde{v}_{-}}\right) ^{2/\sqrt{3}}I\right] ,
\end{equation}
where $I$ is the integral
\begin{equation}
I=\int_{c_{s}}^{\xi _{0}}\left( \frac{1+\xi }{1-\xi }\right) ^{2/\sqrt{3}}
\frac{\left( \xi -c_{s}\right) ^{2}}{1-\xi ^{2}}d\xi .
\end{equation}
The change of variable $x=\left( 1+\xi \right) /\left( 1-\xi \right) $ leads
to the simpler expression
\begin{equation}
I=\int \frac{1}{2}x^{2/\sqrt{3}-1}\left( 1-c_{s}-\frac{2}{x+1}\right) ^{2}dx.
\end{equation}
This integral can be expressed in terms of the hypergeometric
function $_{2}F_{1}$ \cite{grads}. We obtain
\begin{equation}
I=\frac{1}{2}\left[ f\left( \xi _{0}\right) -f\left( c_{s}\right) \right] ,
\end{equation}
where\footnote{See  Eqs. 3.194-1, 9.131-1 and 9.137-2 of Ref.
\cite{grads}.}
\begin{equation}
f\left( \xi \right) =\left( \frac{1+\xi }{1-\xi }\right) ^{\frac{2}{\sqrt{3}}}
\left\{ \frac{2}{\sqrt{3}}-1+\left( 1-\xi \right) \left[ 2-\, _{2}F_{1}(1,1,
\frac{2}{\sqrt{3}}+1,\frac{1+\xi }{2})\right] \right\} .
\end{equation}

\subsection{Subsonic deflagrations}

The profile for subsonic deflagrations is very simple in the planar
case. The fluid velocity is a constant $v\equiv
\tilde{v}_{+}=\tilde{v}_{1}$ between $\xi _{w}$ and $\xi
_{\mathrm{sh} } $, and vanishes outside that region. Thus, we have
$w_{+}=w_{1}$ and $\alpha _{+}=\alpha _{1}$, and Eqs.
(\ref{alfamalfa1}-\ref{alfamadefla}) give an algebraic equation for
$\tilde{v}_{1}$. The equation is simpler in terms of $
\xi_{\mathrm{sh}}$,
\begin{equation}
\left( 3\xi _{\mathrm{sh}}^{2}-1\right) ^{2}+\xi _{\mathrm{sh}}
\left( 3\xi _{\mathrm{sh}}^{2}-1\right) \frac{1-3\xi _{w}^{2}}{\xi _{w}}=
\frac{9}{2}\alpha
_{N}/\gamma _{\mathrm{sh}}^{4}.
\end{equation}
Solving  for $\xi _{\mathrm{sh}}$ as a function of $\alpha _{N}$ and
$v_{w}$ amounts to finding the roots of a quartic polynomial. The
algebraic expressions for the solutions are quite cumbersome and we
shall not write them down. Only one of the four solutions gives $\xi
_{\mathrm{sh}}\geq c_{s}$. The integral in Eq. (\ref{kappa}) is
trivial since $v$ is a constant, and the efficiency factor is given
by
\begin{equation}
\kappa =\frac{1}{\xi _{w}}\frac{w_{1}}{\varepsilon }\tilde{v}_{1}^{2}
\tilde{\gamma}_{1}^{2}\left( \xi _{\mathrm{sh}}-\xi _{w}\right) ,
\end{equation}
where $\tilde{v}_{1}$ is given by Eq. (\ref{v1t}) as a function of
$\xi _{\mathrm{sh}}$, and $w_{1}/\varepsilon $ is given by Eq.
(\ref{w1}) as a function of $\xi _{w}$ and $\alpha _{N}$.

\subsection{Supersonic deflagrations}

In this case the shock wave in front of the wall is similar to that
of subsonic deflagrations. We have again $
\tilde{v}_{+}=\tilde{v}_{1}$, $ w_{+}=w_{1} $, and $\alpha
_{+}=\alpha _{1}$. From Eqs. (\ref{alfa1xish}), (\ref{alfamahibri}),
and (\ref{v1t}) we obtain
\begin{equation}
\gamma _{w}^{2}\left[ \xi _{\mathrm{sh}}(1-\sqrt{3}\xi _{w})-
\frac{3\xi _{\mathrm{sh}}^{2}-1}{2}(\xi _{w}-\sqrt{3})\right] ^{2}=
\frac{9}{4}\frac{\alpha _{N}}{\gamma _{\mathrm{sh}}^{4}}, \label{eqhib}
\end{equation}
which trivially reduces to a quadratic equation. The solution is
\begin{equation}
\xi _{\mathrm{sh}}=\sqrt{\frac{1}{3}+\frac{2\frac{\sqrt{\alpha _{N}}}{\gamma
_{w}}\ x+\left( \xi _{w}-\frac{1}{\sqrt{3}}\right) ^{2}}{3\ x^{2}}}+\frac{%
\xi _{w}-\frac{1}{\sqrt{3}}}{\sqrt{3}\ x} \label{xishhibpla}
\end{equation}
where $x=\sqrt{3}-\xi _{w}+\sqrt{\alpha _{N}}/\gamma _{w}$. The
rarefaction wave behind the wall is given by the solution
$v_{\mathrm{rar}}\left( \xi \right) $. In the Jouguet case,
$\tilde{v}_{-}$ is given by Eq. (\ref{vmethib}), which implies that
$\xi _{0}=\xi _{w}$. The efficiency factor is given by
\begin{equation}
\kappa =\frac{w_{-}}{\xi _{w}\varepsilon }\frac{3}{4}\left(
\frac{1-\xi _{w}}{1+\xi _{w}}\right)
^{\frac{2}{\sqrt{3}}}\left[ f\left( \xi _{w}\right)
-f\left( c_{s}\right) \right] +\frac{w_{1}}{\xi _{w}\varepsilon }
\tilde{v}_{1}^{2}\tilde{\gamma}_{1}^{2}
\left( \xi _{\mathrm{sh}}-\xi _{w}\right) .
\end{equation}
The value of $w_{-}$ in the Jouguet deflagration case is given by Eq.
(\ref{wmehib}) as a function of $w_{+}=w_{1}$, and depends on
$v_{+}=\left( \tilde{v}_{1}-\xi _{w}\right) /\left(
1-\tilde{v}_{1}\xi _{w}\right) $, with $\tilde{v}_{1}$ given by Eq.
(\ref{v1t}). The value of $w_1/\varepsilon$ is again  given by Eq.
(\ref{w1}) as a function of $\xi _{w}$ and $\alpha _{N}$.

For $\xi_w=v^{\mathrm{det}}_J(\alpha_N)$, Eq. (\ref{xishhibpla})
gives $\xi_{\mathrm{sh}}=\xi_w$ [this is more easily checked by
setting $\xi_{\mathrm{sh}}=\xi_w$ in Eq. (\ref{eqhib})]. Hence, at
the Jouguet detonation velocity, the shock disappears and the profile
for the supersonic deflagration matches the profile for the
detonation.

\section{Gravitational waves from real detonations and
deflagrations} \label{gw}

In section \ref{fluid} we have discussed the disturbance that phase
transition fronts  cause on the plasma. In the present section we
consider a particular consequence of such disturbance, namely, the
generation of gravity waves. As we have already mentioned,
gravitational radiation can only be produced once bubbles collide and
lose their spherical symmetry. In fact, the ``bubble collisions''
mechanism \cite{kt93,kkt94,cds08,hk08} is based on the envelope
approximation \cite{kt93}, which consists of taking into account only
the motion of the uncollided walls. The thickness of the shell in
which the energy of the fluid is concentrated is relevant for this
mechanism, and the envelope approximation assumes that the energy
concentrations are infinitesimally thin. On the other hand, at a
cosmological phase transition the Reynolds number is large enough for
bubble collisions to cause the onset of turbulence  \cite{kkt94}.
Turbulence turns out to be a more effective source of gravitational
radiation than bubble collisions \cite{kkt94,gw,cd06}. In an
electrically conducting fluid and in the presence of magnetic fields,
turbulence develops in a completely different way. This gives a third
mechanism for generation of GWs in a phase transition (see e.g.
\cite{cd06,kgr08}).

The energy density of gravitational radiation is usually expressed in
terms of the quantity
\begin{equation}
h^2\Omega _{GW}\left( f\right) =\frac{h^2}{\rho _{c}}\frac{d\rho
_{GW}}{d\log f},
\end{equation}
where $\rho _{GW}$ is the energy density of the GWs, $f$ is the
frequency, and $\rho _{c} $ is the critical energy density today,
$\rho _{c}=3H_{0}^{2}/8\pi G$, with $H_0=100\, h\, \mathrm{ km \,
s}^{-1} \mathrm{Mpc}^{-1}$. The GW spectrum depends on the details of
the phase transition and on the generating mechanism. Nevertheless,
the peak frequency $f_p$ is generally determined by the typical
length scale of the source. In a first-order phase transition, the
latter is the bubble size $d$, which is proportional to the duration
$\Delta t$ of the phase transition, $d\sim v_w\Delta t$. The time
$\Delta t$ is in turn a fraction of the Hubble time. Once redshifted
to today, the peak frequency is roughly given by
\begin{equation}
f_{p}\sim 10^{-2}\mathrm{mHz}\frac{H_{\ast}^{-1}}{\Delta t}
\frac{T_{\ast }}{100\mathrm{GeV}}
\end{equation}
where $H_{\ast }$ and $T_{\ast }$ are the Hubble rate and the
temperature at the moment of the phase transition. The sensitivity
peak of the space interferometer LISA is expected to be $h^{2}\Omega
\sim 10^{-12}$ at a  frequency $f\sim 1\mathrm{mHz}$. Quite
interestingly, GWs produced at the temperature scale of the
electroweak phase transition, $T_{\ast}\sim 100GeV$, will have a
characteristic frequency around $f_p\sim 1\mathrm{mHz}$ for $\Delta t
\sim 10^{-2} H_{\ast}^{-1}$, which is a possible value for the
duration of the phase transition. This motivated the investigation of
the GW signal from the electroweak phase transition
\cite{ms10,gwew,amnr02}.

For simplicity, we shall consider only GWs from turbulence, for which
we shall use the analytic approximation obtained in Ref. \cite{cd06}
(we have checked, using the fit given in Ref. \cite{hk08}, that the
intensity resulting from bubble collisions is an order of magnitude
smaller). The GW energy density $\Omega_p$ at the peak frequency
depends on the length scale $d\sim v_w\Delta t$. Thus, we have
\begin{equation}
\Omega _{p}\approx \frac{9\Omega _{R}}{32\pi }v_{w}^{2}
\left(\frac{\Delta t}{H_{\ast}^{-1}}\right) ^{2}
\kappa^2\alpha_N^2
\left\{ \begin{array}{ccc}
                4 & \mathrm{for} & v\leq 1/2, \\
                1/v^2 & \mathrm{for} & v\geq 1/2, \\
              \end{array}
 \right. \label{omp}
\end{equation}
where $\Omega _{R}\approx  5\times 10^{-5}$ is the radiation energy
density parameter $\Omega _{R}=\rho_R/\rho_c$ today, and $v$ is the
characteristic eddy velocity, defined by
$v^2=\frac{3}{2}\kappa\alpha_N$.  Equation (\ref{omp}) should be
valid both for detonations and deflagrations, although the time
$\Delta t$ must be calculated differently in each case.

Notice  that $\Omega_p$ is proportional to $v_w^2$ and to $\kappa^2$.
The wall velocity is determined by hydrodynamics and by
\emph{microphysics} (see, e.g., \cite{ikkl94}). As a consequence,
$v_w$ depends on a friction parameter $\eta$ as well as on the
nucleation temperature $T_N$. Different approximations have been used
for the friction force  (see, e.g., \cite{ms09,ekns10}). For
generality, we shall leave the result expressed in terms of $v_w$. We
do not expect $v_w$ to depend significantly on the geometry. In the
case of deflagrations, the dependence of $v_w$ on $T_N$ is affected
by the shock wave, which depends on the wall geometry. In the case of
detonations, $v_w$ is completely determined by the discontinuity
equations (\ref{landau2}) and the friction, and does not depend on
the geometry at all. In any case, as explained before, there is no
reason to assume any particular symmetry after bubbles collide, and
we shall use the planar wall results.

\begin{figure}[hbt] \centering \epsfxsize=10cm \leavevmode
\epsfbox{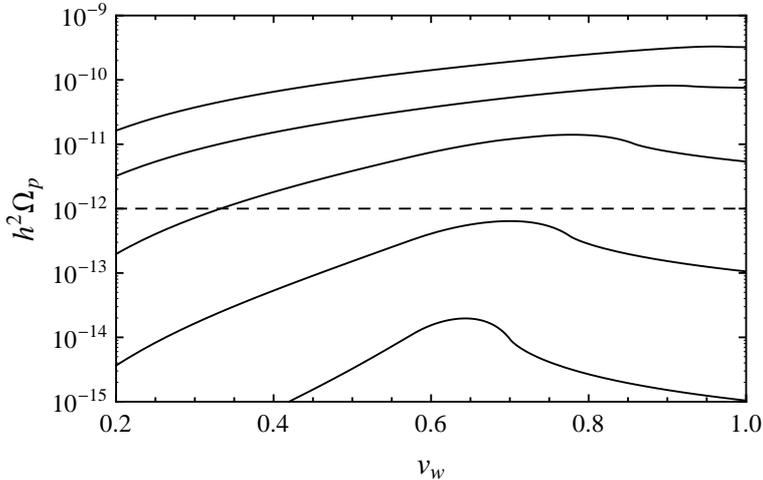} \caption{The intensity of gravitational radiation
from turbulence at a phase transition with $\Delta t H_{\ast}=10^{-2}$
and $T_*=100$GeV,
for $\alpha_N=0.03$, 0.1, 0.3, 1, and 3 (from bottom to top). The dashed
line corresponds to the peak sensitivity of LISA.}
\label{figomvw}
\end{figure}

In Fig. \ref{figomvw}  we plot the peak amplitude $ \Omega _{p}$ of
GWs from turbulence at the electroweak phase transition for some
values of $\alpha _{N}$, as a function of the wall velocity $v_{w}$.
We chose $\Delta t H_{\ast }=10^{-2}$, which gives $f_{p}\sim 1mHz$.
We see that the phase transition needs not be too strong, i.e., with
$\alpha_N\gtrsim 0.3$ we obtain intensities above the peak
sensitivity of LISA. This is important, since the value of $\alpha_N$
was  found to be $\alpha_N\lesssim 1$ for several models of the
electroweak phase transition  \cite{amnr02}. For $\alpha_N\sim 1$ we
get $h^2\Omega_p$ as large as $\sim 10^{-9}$.

\begin{figure}[hbt] \centering \epsfysize=6cm \leavevmode
\epsfbox{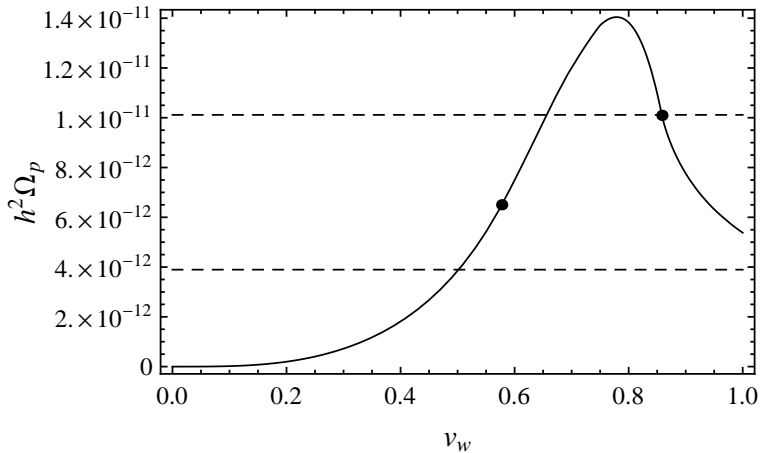} \caption{The same as Fig. \ref{figomvw},
for $\alpha_N=0.3$ alone. The dashed lines indicate the results
obtained from the correct (upper) and the wrong (lower) values
of $\kappa _{J}\left( \alpha _{N}\right) $. The two points
indicate the cases $v_w=c_s$ and $v_w=v_J^{\mathrm{det}}(\alpha_N)$,
which separate subsonic deflagrations, supersonic deflagrations,
and detonations.} \label{figjougweak}
\end{figure}

Since  only Jouguet detonations are in general considered in the
literature on GWs, the efficiency factor is in principle
overestimated with respect to real detonations. However, as pointed
out in Ref. \cite{ekns10}, the value of the Jouguet detonation
efficiency factor $\kappa _{J}\left( \alpha _{N}\right) $ is
underestimated in the literature, due to a missing factor $v_{w}^{3}$
in the original paper \cite{kkt94}. This mistake compensates the fact
that $\kappa _{J}\left( \alpha _{N}\right) $ is larger than the
efficiency factor $\kappa(\alpha_N,v_w)$ for weak detonations. The
effect of this compensation on the intensity of GWs is shown in Fig.
\ref{figjougweak}, where we plotted $\Omega_p$ as a function of $v_w$
for $\alpha _{N}=0.3$, together with the correct (upper dashed line)
and the wrong (lower dashed line) values for the Jouguet detonation
case. The efficiency factor for weak detonations lies between these
two values. Notice, also, that supersonic deflagrations give the
largest GW amplitudes, and that even  subsonic deflagrations can give
intensities comparable to those of detonations.

\section{Conclusions} \label{conclu}

We have studied the motion of phase transition fronts in a
first-order cosmological phase transition, focusing on the energy
injected into bulk motions of the plasma, which is a relevant
quantity for the generation of cosmological relics such as
gravitational waves. This issue was recently addressed in Ref.
\cite{ekns10} for the case of spherical bubbles. However, the GWs are
generated once the bubbles (or the shock fronts) collide, so that the
spherical symmetry is lost. Therefore, any bubble symmetry one may
assume will be just an approximation. In order to study the
dependence  on the wall geometry, we have considered bubble walls
with spherical, cylindrical, and plane symmetry, for all the possible
hydrodynamic propagation modes, namely, subsonic deflagrations,
supersonic deflagrations, and detonations.

We have seen that the  region around the wall in which the energy
spreads  can be rather different for each wall geometry. In
particular, for planar walls the region is larger, since the energy
spreads in only one direction. For the strongest processes allowed,
i.e., Jouguet deflagrations and Jouguet detonations, the kinetic
energy of the fluid is, for the three geometries, concentrated in a
thin region around the wall. This is because these processes produce
a stronger disturbance of the plasma than weak processes; thus, the
injected energy is larger and, hence, more difficult to distribute.

The efficiency factor $\kappa$, i.e., the part of the injected energy
which goes into bulk motions (relative to the released vacuum
energy), has a rather weak dependence on the wall geometry. This is
an important result, since the walls can take arbitrary forms after
colliding. The dependence on the wall geometry is stronger for small
wall velocities, $\xi_w \lesssim 0.2$. For small velocities, however,
the efficiency factor is small and will not play a relevant role in
the cosmological consequences of the phase transition. Thus, it is
clear that, for applications, it is convenient to consider planar
walls, for which we have obtained exact analytical expressions for
$\kappa$ (alternatively, one can use the numerical fits given in Ref.
\cite{ekns10} for the spherical case).

The efficiency factor peaks for supersonic deflagrations, which are
Jouguet processes. Thus, $\kappa$ is in general sizeable for fast
(close to the speed of sound) subsonic deflagrations, and for the
slowest detonations, i.e., those solutions which are close to the
Jouguet velocity. On the other hand, $\kappa$ can decrease
considerably for fast detonations, and vanishes for $v_w\to 0$. We
have applied the results for the planar case to the estimation of the
gravitational wave signal from turbulence at the electroweak phase
transition. The GW amplitude peaks for supersonic deflagrations. It
is interesting to notice that subsonic deflagrations can give
efficiency factors larger than those given by detonations. Although
the amplitude of the gravity waves may depend (according to the
generation mechanism) on the wall velocity as well as on the
efficiency factor, we have seen that subsonic deflagrations can
produce GWs of intensity comparable to that of detonations.

\end{document}